\documentclass[pre,aps,twocolumn, superscriptaddress]{revtex4}
\usepackage{amsmath}
\usepackage{epsfig}
\usepackage{amssymb}
\usepackage{color}

\begin{document}

\title{Thermodynamics of a Brownian particle in a non-confining
  potential}

\author{Oded Farago} 
\affiliation{Department of Biomedical
Engineering, Ben-Gurion University of the Negev, Be'er Sheva 85105,
Israel} 

\begin{abstract}

We consider the overdamped Brownian dynamics of a particle starting
inside a square potential well which, upon exiting the well,
experiences a flat potential where it is free to diffuse. We calculate
the particle's probability distribution function (PDF) at coordinate
$x$ and time $t$, $P(x,t)$, by solving the corresponding Smoluchowski
equation. The solution is expressed by a multipole expansion, with
each term decaying $t^{1/2}$ faster than the previous one. At
asymptotically large times, the PDF {\em outside}\/ the well converges
to the Gaussian PDF of a free Brownian particle. The average energy,
which is proportional to the probability of finding the particle {\em
  inside}\/ the well, diminishes as $E\sim 1/t^{1/2}$. Interestingly,
we find that the free energy of the particle, $F$, approaches the free
energy of a freely diffusing particle, $F_0$, as $\delta F=F-F_0\sim
1/t$, i.e., at a rate faster than $E$. We provide analytical and
computational evidences that this scaling behavior of $\delta F$ is a
general feature of Brownian dynamics in non-confining potential
fields. Furthermore, we argue that $\delta F$ represents a diminishing
entropic component which is localized in the region of the potential,
and which diffuses away with the spreading particle without being
transferred to the heat bath.

\end{abstract} 

\maketitle

\section{Introduction}

Single particle Brownian motion constitutes one of the fundamental
models in statistical-mechanics. It is the simplest example of
diffusion, which is one of the most important mechanisms of molecular
and energy transport in nature~\cite{crank_book}. It is also
used as a mean to introduce the elusive concept of coupling between a
thermodynamic system and a heat bath, which forms the basis for
Molecular Dynamics simulations~\cite{frenkel_smit}. When the particle
is confined to a finite space by an external potential field, $U(x)$,
its probability distribution function (PDF) relaxes at large times to
the Boltzmann's equilibrium distribution:
$P(x,t\rightarrow\infty)=P_{\rm eq}(x)=\exp[-\beta
  U(x)]/Z$~\cite{pathria}. Here, $\beta=1/k_BT$, where $k_B$ is
Boltzmann's constant, $T$ is the temperature which is assumed to be
uniform in space, and $Z=\int_{-\infty}^{\infty}\exp[-\beta U(x)]dx$
is the normalizing partition function. Brownian dynamics in confined
(closed) molecular systems are perceived as stochastic trajectories in
the configurational phase space~\cite{chen_04}. For a single Brownian
particle it is expected, by virtue of the ergodicity hypothesis, that
the time average of an observable over a sufficiently long stochastic
trajectory coincides with the ensemble average over the equilibrium
PDF, $P_{\rm eq}(x)$~\cite{keller}.

A very different scenario arises when the particle diffuses in a
spatially unbounded system. Consider, for instance, an overdamped
Brownian particle moving in a potential field which has the form
\begin{equation}
 U(x) = \left\{ \begin{array}{ll}
         U(x) & \mbox{{\rm for}  $|x| < x_0$}\\
         0 & \mbox{{\rm for} $|x|\geq x_0$},
\end{array} \right.
\label{eq:potential}
\end{equation}
or, more generally, a potential field that decays faster than
$1/x$. The PDF of the particle,
$P(x,t)$, solves the Smoluchowski equation~\cite{smoluch}
\begin{equation}
  \frac{\partial P(x,t)}{\partial t}=D\frac{\partial}{\partial
    x}\left\{e^{-\beta U(x)}\frac{\partial}{\partial x}
    \left[e^{\beta U(x)}P(x,t)\right]\right\},
  \label{eq:smoluch}
\end{equation}
 where $D$ is the diffusion coefficient of the medium. In a
 non-confining potential field of the form of
 Eq.~(\ref{eq:potential}), the partition function $Z$ is
 divergent. The PDF does not relax to the Boltzmann equilibrium
 distribution but rather continues to spread indefinitely.  Is it
 still possible to define a statistical-mechanical framework for such
 a class of non-equilibrium processes? This question has been recently
 addressed by Aghion et al.~who argued that the long-time asymptotic
 form of the PDF is given by~\cite{kb1,kb2}
 \begin{equation}
   P(x,t)\simeq \frac{e^{-\beta U(x)}}{\sqrt{4\pi
       Dt}}e^{-x^2/4Dt}=e^{-\beta U(x)}G(x,t),
   \label{eq:barkai}
 \end{equation}
 where $G(x,t)=\exp(-x^2/4Dt)/\sqrt{4\pi Dt}$ is the ``fundamental''
 Gaussian solution, i.e., the PDF of a particle diffusing in a flat
 potential (free diffusion), subject to the Dirac delta-function
 initial condition, $P(x,t=0)=\delta(x)$. Thus, for $x\ll\sqrt{Dt}$,
 we have $P(x,t)\simeq \exp[-\beta U(x)]/\sqrt{4\pi Dt}$, which has a
 similar form to the Boltzmann equilibrium PDF, but with a
 time-dependent partition coefficient
\begin{equation}
  Z^*=\sqrt{4\pi Dt}.
  \label{eq:partcoef}
\end{equation}
Writing that $\lim_{t\rightarrow\infty} Z^*P(x,t)=\exp[-\beta U(x)]$,
means that the Boltzmann factor is reached at sufficiently long times
and plays the role of an infinite invariant
density~\cite{kb1,kb2,infden1,dechant11,infden2,wang19,infden3,infden4,infden5}.
This paves the way to formulating a non-equilibrium statistical
framework which is based on concepts from the infinite ergodic theory
relating ensemble and time averages of non-normalizable densities.

From Eq.~(\ref{eq:barkai}) it follows that for $|x|\geq x_0$ [outside
  the non-confining potential (\ref{eq:potential})] at large times,
$P(x,t)\simeq G(x,t)$~\cite{comment1}. That the PDF, $P(x,t)$,
converges to the form of the fundamental Gaussian PDF, $G(x,t)$, means
that, in a sense, the latter plays here a role reminiscent of the
equilibrium Boltzmann distribution in a closed system (see
footnote~\cite{kb3}). It is, therefore, interesting to check how
different thermodynamic quantities approach the values of their
counterparts in the free diffusion [$U(x)=0$] case. The energy and
entropy of the a freely diffusing particle are given by:
\begin{eqnarray}
  E_0&=&0 \label{eq:e0}\\
  S_0&=&-k_B\int_{-\infty}^{\infty}
  G(x,t)\ln[G(x,t)]dx\nonumber \\
  &=&k_B\left[\ln(Z^*)+\frac{1}{2}\right]
\label{eq:s0}
\end{eqnarray}
From Eqs.~(\ref{eq:potential}) and (\ref{eq:barkai}) it is easy to see
that for $U(x)\neq 0$, the excess energy of the particle, $\delta
E=E-E_0\simeq\int_{-\infty}^{\infty}U(x)\exp[-\beta U(x)]dx/Z^*$,
converges to zero with time as $\sim 1/t^{1/2}$. Similarly, one can
check that the excess entropy $\delta
S=S-S_0=-k_B\int_{-\infty}^{\infty} P(x,t)\ln[P(x,t)]dx-S_0$, also
scales $\sim 1/t^{1/2}$.

With that said, it is important to understand that the PDF
(\ref{eq:barkai}) is {\em not}\/ a solution of Eq.~(\ref{eq:smoluch}),
but rather the asymptotic form of the solution at large times (see
comment ref.~\cite{comment1}). Eq.~(\ref{eq:barkai}) is, in fact, the
first (leading) in a series of terms, each of which decaying at large
times $t^{1/2}$ faster than the previous one. In ref.~\cite{kb2}, the
first correction term to Eq.~(\ref{eq:barkai}) was calculated using
eigenfunction expansion. Here, we focus on a specific example of a
square potential well, $U(x)=-U$ in Eq.~(\ref{eq:potential}). For this
example, we calculate the first two correction terms to
Eq.~(\ref{eq:barkai}), which are sufficient for characterizing the
asymptotic thermodynamic behavior of the system. This is done by using
the method of images, taking advantage of the fact that for the
derivation of the first two correction terms in the solution series
expansion (in powers of $1/t^{1/2}$), only two images are needed. We
find that while the excess energy and entropy with respect to free
diffusion diminishes $\sim 1/t^{1/2}$ (see above), the excess
Helmholtz free energy decays faster: $\delta F=\delta E-T\delta S\sim
1/t$. The square well example is studied in
section~\ref{sec:square}. In section~\ref{sec:general} we generalize
the discussion to an arbitrary non-confining potential field and find
that $\delta F/k_BT=A^2/2Dt$, where $A$ is a constant with
dimensionality of length that can be related the second virial
coefficient of the potential. This result constitutes a new
thermodynamic relation for the overdamped evolution of Brownian
particles in non-confining potentials. It is discussed in
section~\ref{sec:summary}, where we argue that $\delta F$ represents a
diminishing component of the entropy which is localized in the region
of the potential, and which is lost when the particle diffuses away
from the potential well.

\section{The case of a square potential well}
\label{sec:square}

\subsection{The spatial distribution}

For a square potential well $U(x)=-U$, the solutions both inside
($|x|<x_0$), $P_<(x,t)$, and outside ($|x|>x_0$), $P_>(x,t)$, the well
satisfy the the free diffusion equation
$\partial_tP=D\partial_{xx}P$. They must be matched by two boundary
conditions (BCs) at $x_0$. The first one is, obviously, the continuity
of the flux
\begin{equation}
  \partial_x P_<(x_0,t)=\partial_x P_>(x_0,t).
  \label{eq:bcflux}
\end{equation}
The second BC, which is known as the ``imperfect contact''
condition~\cite{carr16}, reads
\begin{equation}
  e^{-\beta U}P_<(x_0,t)= P_>(x_0,t)
  \label{eq:bcjump}
\end{equation}
 This condition is widely used in many theoretical studies of mass and
 heat diffusion problems across sharp
 interfaces~\cite{korabel11,pino16,sheils17} (see a brief explanation
 and derivation in footnote~\cite{comment2}). The coefficient
\begin{equation}
  \sigma=e^{-\beta U}
  \label{eq:sigma}
\end{equation}
is called the partition coefficient of the interface.

The problem of diffusion from a square-well can be solved using the
method of images. An ``image'' particle of size $q$ located at $x=a$
generates a Gaussian distribution
\begin{widetext}
\begin{equation}
  P_{\rm image}(x,t)=qG(x-a,t)=q\frac{e^{-(x-a)^2/4Dt}}{\sqrt{4\pi
      Dt}}=q\frac{e^{-x^2/4Dt}}{\sqrt{4\pi
      Dt}}\left[1+\frac{xa}{2Dt}+\frac{a^2}{4Dt}\left(\frac{x^2}{2Dt}-1\right)
    +\cdots\right].
  \label{eq:image}
\end{equation}
\end{widetext}
We note that $P_{\rm image}$ is {\em not}\/ normalized to unity
[$\int_{-\infty}^{\infty}P_{\rm image}(x,t)dx=q$].  We also note
that, up to a multiplicative constant, the $n$-th term
($n=0,1,2,\ldots$) in this expansion of $G(x-a,t)$ has the form
\begin{equation}
  P_n(x,t)=G(x,t)\left(\frac{a}{\sqrt{2Dt}}\right)^nH_n
  \left(\frac{x}{\sqrt{2Dt}}\right),
  \label{eq:hermite}
\end{equation}
where $H_n$ is the $n$-th probabilists' Hermite polynomial
[$H_0(y)=1$; $H_1(y)=y$; $H_2(y)=y^2-1$; $H_3(y)=y^3-3y$;
  $\ldots$]. Eq.~(\ref{eq:image}) is essentially a multipole
expansion. The leading term, $P_0$ , is the fundamental solution of a
``monopole'', namely a Brownian particle starting at the origin. The
next term ($n=1$) describes the PDF of a dipole, i.e., two opposite
images located symmetrically with respect to the origin. Then the
following terms correspond to a linear quadrupole setting ($n=2$),
octupole ($n=3$), etc. From the linearity of the free diffusion
equation it follows that each function $P_n(x,t)$ (\ref{eq:hermite})
is itself a solution of this equation. Therefore, a linear combination
of $P_n(x,t)$
\begin{equation}
  P(x,t)=G(x,t)\sum_{n=0}^{\infty}\left(\frac{c_n}{\sqrt{2Dt}}\right)^nH_n
  \left(\frac{x}{\sqrt{2Dt}}\right),
  \label{eq:hermite2}
\end{equation}
where $c_n$ are constants with dimenstionlity of length, is also a
solution of the free diffusion equation has the form~\cite{choi20}.

\begin{figure}[t]
\centering\includegraphics[width=0.45\textwidth]{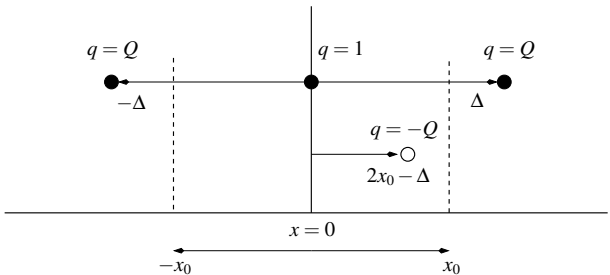}
\caption{A schematic explaining the solution by the method of
  images. The diffusing particle is represented by the solid circle at
  the origin and has a size $q=1$. In order to derive the PDF inside
  the well ($x<x_0$), we place two images of size $q=Q$ at
  $x=\pm\Delta$, where $\Delta>x_0$. These images are also represented
  by solid circles. For the calculation of the PDF in the region to
  the right of the well ($x>x_0$), we replace the image at $x=\Delta$
  with an opposite image of size $-Q$ which is placed at $2x_0-\Delta$
  (depicted with an open circle).}
\label{fig0}
\end{figure}

With the above in mind, we return to the escape problem from the
square well, subject to delta function initial conditions
$P(x,0)=\delta(x)$. We note the following: (i) Because of the symmetry
of the problem with respect to reflection around the origin, we must
have that $P(x,t)=P(-x,t)$, which means that we only need to solve the
PDF for $x>0$. Symmetry also implies that the PDF inside the well,
$P_<(x,t)$, is an even function and, thus, when expressed as in
Eq.~(\ref{eq:hermite2}), it contains only the even terms. This ensures
that the probability flux at the origin vanishes
[$\partial_xP_<(0,t)=0$]. (ii) As we will see later in
section~\ref{sec:general}, the asymptotic thermodynamic behavior is
captured by the terms up to order $1/t^{3/2}$, which means that we
only need to calculate the first three moments in
Eq.~(\ref{eq:hermite2}) or, equivalently, express the PDF as the sum
of the PDFs of three particles. These are located as shown in
fig.~\ref{fig0}. The central particle of size $q=1$, which is located
at the origin, represents the diffusing Brownian particle. Since
$P_<(x,t)$ has no dipole term, we place two image particles of size
$q=Q$ at $\pm\Delta$. We require that $\Delta>x_0$, i.e., put these
images outside the potential well in order to guarantee that the
delta-function initial condition is satisfied (even though we are
interested in the behavior at asymptotically large times). For
$P_>(x,t)$, we keep the particles at $x=0$ and $x=-\Delta$, and
replace the image at $\Delta>x_0$ with an opposite image of size $-Q$
located symmetrically with respect to the boundary at $x_0$, i.e., at
$x=x_0-(\Delta-x_0)=2x_0-\Delta$ (depicted with an open circle in
fig.~\ref{fig0}). With this replacement we accomplish two things:
First, the fact $P_>(x,t)$ is represented as the sum of the PDFs of
three particles, none of which is located at $x>x_0$, ensures that the
delta-function initial boundary condition is satisfied. Second, the
exchange of the image $Q$ with an opposite image $-Q$ locate
symmetrically with respect to the boundary ensures that the flux is
continuous at $x=x_0$, which means that BC (\ref{eq:bcflux}) is
satisfied.

The values of $Q$ and $\Delta$ can be now found by writing both
$P_<(x,t)$ and $P_>(x,t)$ in the form of Eq.~(\ref{eq:hermite2}), and
imposing BC (\ref{eq:bcjump}) [with the partition coefficient $\sigma$
  defined in Eq.~(\ref{eq:sigma})] to order $1/t^{3/2}$. Comparing the
terms proportional to $1/t^{1/2}$ yields $Q=(1-\sigma)/(2\sigma)$, and
from the terms proportional to $1/t^{3/2}$ we find that
$\Delta=2x_0/\sigma$. Since we demand that $\Delta>x_0$, we must
restrict the discussion in what follows to $0<\sigma<2$. Notice that
for $\sigma>1$, we consider a potential step rather than a potential
well. With the above values of $Q$ and $\Delta$, the PDFs are given by
\begin{widetext}
\begin{eqnarray}
  P_<(x,t)\!&=&G(x,t)
  \left[\frac{1}{\sigma}-\left(\frac{1-\sigma}{\sigma^3}\right)
    \frac{x_0^2}{Dt}+{\cal O}\left(\frac{1}{t^2}\right)\right]
      \label{eq:p1small}\\
      P_>(x,t)\!&=&G(x,t)
      \left[1-\frac{1-\sigma}{\sigma}\frac{xx_0}{2Dt}
        +\frac{(1-\sigma)(2-\sigma)}{\sigma^2}\frac{x_0^2}{2Dt}
      \left(\frac{x^2}{2Dt}-1\right)+{\cal
        O}\left(\frac{1}{t^{3/2}}\right)\right].
  \label{eq:p1large}
\end{eqnarray}
\end{widetext}
The scaling behavior of $x$ with $t$ is $x\sim x_0\sim t^0$ in $P_<$,
and $x\sim \sqrt{Dt}\sim t^{1/2}$ in $P_>$. Also recall that
$G(x,t)\sim 1/t^{1/2}$. Thus, in Eq.~(\ref{eq:p1large}) the terms
scale as $1/t^{1/2}$, $1/t$, $1/t^{3/2} \ldots$, while in
Eq.~(\ref{eq:p1small}) the terms with scaling $\sim 1/t^n$, where $n$
is an integer, are missing because $P_<$ is even. If we keep only the
leading terms in Eqs.~(\ref{eq:p1small})-(\ref{eq:p1large}), we get
\begin{eqnarray}
  P_<(x,t)&\simeq&\frac{G(x,t)}{\sigma}
      \label{eq:pleadsmall}\\
      P_>(x,t)&\simeq&G(x,t),
  \label{eq:pleadlarge}
\end{eqnarray}
which is the asymptotic solution Eq.~(\ref{eq:barkai}).

\begin{figure}[t]
\centering\includegraphics[width=0.45\textwidth]{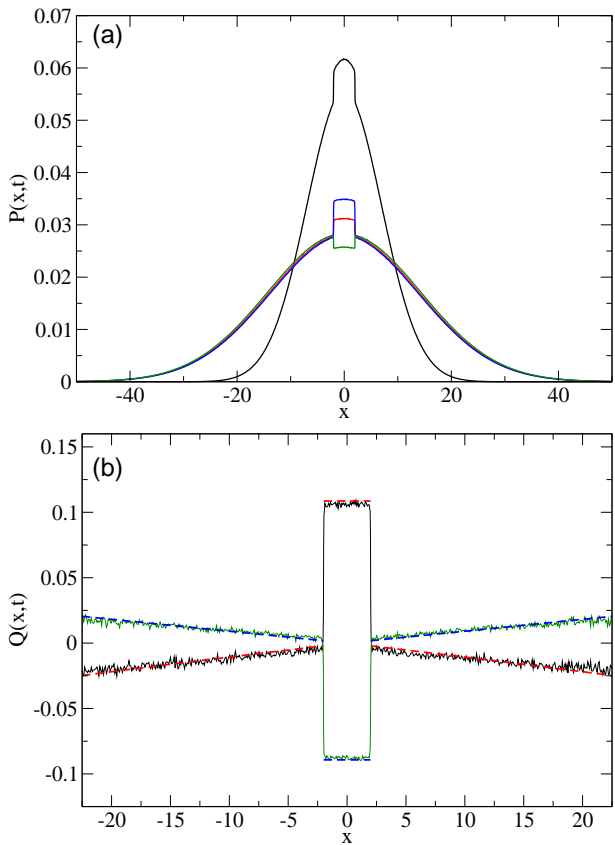}
\caption{(a) The PDF of a particle diffusing in a square well
  potential of size $2x_0=4$ with $D=0.01$. The black curve presents
  the results for $\sigma=0.9$ at $t=2500$. The other curves
  correspond to $t=10^4$ with $\sigma=0.9$ (red), $\sigma=0.8$ (blue),
  and $\sigma=1.1$ (green). (b) The function $Q(x,t)$ (see definition
  in the text) for $\sigma=0.9$ (black) and $\sigma=1.1$ (green) at
  $t=10^4$. Dashed red and blue lines depict the corresponding
  predictions of Eqs.~(\ref{eq:p0small}) and (\ref{eq:p0large}).}
\label{fig1}
\end{figure}

Omitting in Eqs.(\ref{eq:p1small})-(\ref{eq:p1large}) the terms $\sim
1/t^{3/2}$, whose contributions at large times to the PDF is extremely
small and fall below the resolution of the computer simulations, we
write
\begin{eqnarray}
  P_<(x,t)&\simeq&\frac{G(x,t)}{\sigma}
      \label{eq:p0small}\\
      P_>(x,t)&\simeq&G(x,t)
      \left[1-\frac{1-\sigma}{\sigma}\frac{xx_0}{2Dt}\right].
  \label{eq:p0large}
\end{eqnarray}
Fig.~\ref{fig1}(a) shows results for $P(x,t)$ based on $10^8$ Langevin
dynamics trajectories starting at $x=0$, that were generated with the
algorithm presented in ref.~\cite{farago1} (see also
ref.~\cite{farago2}. The algorithm is an extension to discontinuous
potentials of the Gr\o{}nbech-Jensen \& Farago (GJF) integrator for
inertial Langevin dynamics~\cite{gjf}. The friction coefficient in the
simulations is set to $\alpha=k_BT/D$). The system parameters are
$D=0.01$ and $x_0=2$. The black curve shows the PDF for $\sigma=0.9$
at $t=2500$; the other curves correspond to $t=10^4$ with $\sigma=0.9$
(red), $\sigma=0.8$ (blue), and $\sigma=1.1$ (green).  Noticeably, the
last three curves look nearly identical for $x>x_0$, which is
consistent with Eq.~(\ref{eq:p0large}) where the asymptotically
leading term in $P_>(x,t)$ is the fundamental Gaussian solution. To
better test the accuracy of Eqs.~(\ref{eq:p0small}) and
(\ref{eq:p0large}), we plot the function $Q(x,t)\equiv
P(x,t)/G(x,t)-1$ and compare the computational data with the
analytical expressions. This is done in fig.~\ref{fig1}(b), showing
the computational results at $t=10^4$ for $\sigma=0.9$ (black) and
$\sigma=1.1$ (green), along with the corresponding predictions of
Eqs.~(\ref{eq:p0small}) and (\ref{eq:p0large}) (dashed red and blue
lines, respectively). The agreement is, clearly, excellent.  \\ \\
\subsection{The free energy}

In the spirit of the equilibrium canonical ensemble, we define the
time-dependent Helmholtz free energy $F(t)=E(t)-TS(t)$, where the
entropy is given by
\begin{widetext}
\begin{eqnarray}
  S(t)=-k_B\int_{-\infty}^{\infty}P(x,t)\ln P(x,t)dx=
  -2k_B\left\{\int_0^{x_0}P_<(x,t)\ln\left[P_<(x,t)\right]dx
  +\int_{x_0}^{\infty}P_>(x,t)\ln\left[P_>(x,t)\right]\right\},
  \label{eq:entropy}
\end{eqnarray}
\end{widetext}
and the energy
\begin{equation}
  E(t)=\int_{-\infty}^{\infty}U(x)P(x,t)dx=-2U\int_0^{x_0}P_<(x,t)dx.
  \label{eq:energy}
\end{equation}
Inserting expressions (\ref{eq:p0small})-(\ref{eq:p0large}) into
Eqs.~(\ref{eq:entropy})-(\ref{eq:energy}), we arrive after some
calculations at
\begin{eqnarray}
  -TS(t)&=&-TS_0(t)+\frac{2Ux_0}{\sigma Z^*}+{\cal
    O}(1/t)\label{eq:free0}\\
   E(t)&=&E_0-\frac{2Ux_0}{\sigma Z^*}+{\cal
    O}(1/t^{3/2})\label{eq:free1},
\end{eqnarray}
where $Z^*$ is the partition coefficient defined in
(\ref{eq:partcoef}), while $E_0=0$ and $S_0$ are the energy and
entropy of a freely diffusing particle in a flat potential $U=0$ see,
respectively, Eqs.~(\ref{eq:e0}) and (\ref{eq:s0}). From
Eqs.~(\ref{eq:free0})-(\ref{eq:free1}) we conclude that, at large
times, the excess (with respect to free diffusion) energy and entropy
diminishes as $1/Z^*\sim 1/t^{1/2}$, while the excess free energy,
$\delta F=(E-E_0)-T(S-S_0)$, diminishes at a faster rate: $\delta F\sim
1/t$.  From dimensional analysis we can rewrite the above result as
\begin{equation}
  \frac{\delta F}{k_BT} \simeq \frac{A^2}{2Dt}=\frac{A^2}
       {\langle x^2\rangle_{U=0}},
  \label{eq:deltaf1}
\end{equation}
where $A$ is a constant with dimensionality of length, and $\langle
x^2\rangle_{U=0}=2Dt$ is the mean square displacement of a free
particle. In the following section, we consider single particle
diffusion in a general non-confining potential field and show that $A$
is comparable to the (finite) range of the potential well. Another way
to write Eq.~(\ref{eq:deltaf1}) is in the form resembling that of
Einstein's relation $k_BT=D/\mu$, where $\mu$ is the mobility of the
particle. Introducing the time-dependent diffusion coefficient
$\tilde{D}(t)=A^2/2t$, we can write
  \begin{equation}
    \delta F(t)=\frac{\tilde{D}(t)}{\mu},
    \label{eq:deltaf1b}
  \end{equation}
which constitutes a novel linear response (Einstein) relation for
non-confining potentials~\cite{newcomment}.

\section{The general case}
\label{sec:general}

\begin{figure}[tb]
\centering\includegraphics[width=0.45\textwidth]{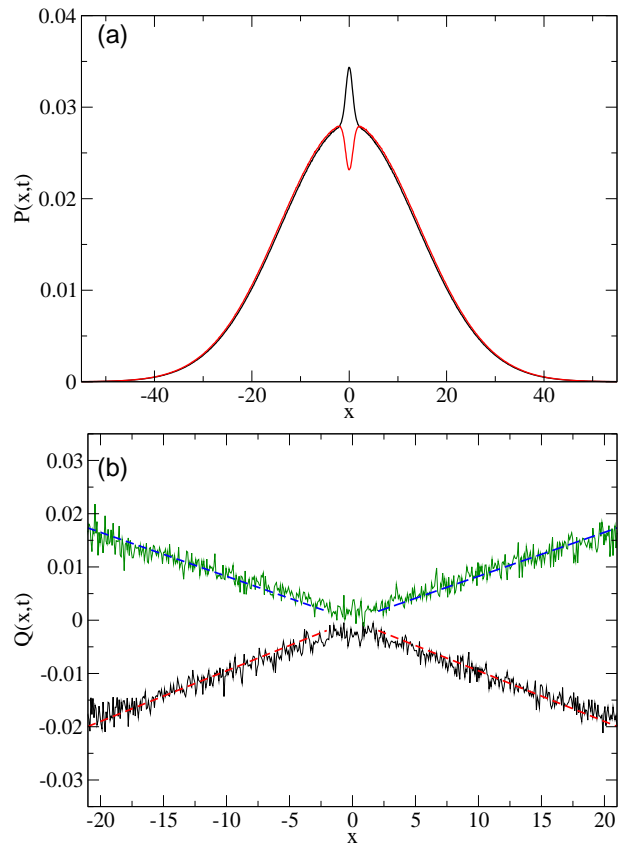}
\caption{(a) The PDF of a particle diffusing ($D=0.01$) in a Gaussian
  potential (\ref{eq:gausspoten}) of size $x_0=1$ with $U=-0.2$
  (black) and $U=+0.2$ (red) at $t=10^4$. (b) The function $Q(x,t)$
  (see definition in the text) for $U=-0.2$ (black) and $U=+0.2$
  (green) at $t=10^4$. Dashed red and blue lines depict the
  corresponding predictions of Eq.~(\ref{eq:general}) with $l_0=0.190$
  and $l_0=-0.165$, respectively.}
\label{fig2}
\end{figure}

For a general non-confining potential $U(x)$, it has been shown in
ref.~\cite{kb2} [cf.~Eq.~(48)] that {\em for large $x$ and $t$}\/, the
PDF is well approximated by
\begin{equation}
  P(x,t)=G\left(1-\frac{l_0|x|}{2Dt}
  +d_2\frac{x^2-2Dt}{4D^2t^2}+\cdots\right),
  \label{eq:general}
\end{equation}
where $G=G(x,t)$ for brevity, 
\begin{equation}
  l_0=\int_{0}^{\infty}\left[e^{-\beta U(x)}-1\right]dx
    \label{eq:l0}
\end{equation}
is related to the second virial coefficient, and $d_2$ is a constant
with dimensionality of $[{\rm length}]^2$. For the square well example
in section~\ref{sec:square}, $l_0=(1-\sigma)x_0/\sigma$, and so the
leading correction (dipole) term in Eq.~(\ref{eq:p1large}) for
$P_>(x,t)$ is nothing but a special case of the corresponding term in
Eq.~(\ref{eq:general}). The constant $d_2$ in the next correction
(quadrupole) term in Eq.~(\ref{eq:general}) depends on the initial
distribution of the particle. In the case of a square well potential
with $P(x,t=0)=\delta(x)$, we have from the comparison of
Eqs.~(\ref{eq:general}) and (\ref{eq:p1large}) that
$d_2=(1-\sigma)(2-\sigma)x_0^2=2l_0^2+l_0x_0$. One may thus speculate
that for the problem of diffusion in a general symmetric potential
$U(x)$ subject to $\delta$-function initial condition at the origin
\begin{equation}
    d_2=2\left\{l_0^2+\int_{0}^{\infty}x\left[e^{-\beta
        U(x)}-1\right]dx\right\}.
       \label{eq:d21}
\end{equation}
Eq.~(\ref{eq:d21}) can be also written as
\begin{equation}
 d_2=2l_0(l_0+l_1),
 \label{eq:d2}
\end{equation}
where the length $l_1$ is defined as
\begin{equation}
l_1=\frac{\int_{0}^{\infty}x\left[e^{-\beta
      U(x)}-1\right]dx}{\int_{0}^{\infty}\left[e^{-\beta
      U(x)}-1\right]dx}
   \label{eq:d22}
\end{equation}
We note the following regarding Eqs.~(\ref{eq:l0})-(\ref{eq:d22}):
\begin{enumerate}
\item The lengths $l_0$ and $l_1$ scale as $t^0$ since the integrands
  in Eqs.~(\ref{eq:l0}) and (\ref{eq:d22}) are non-zero only within
  the limited range of the potential well.
\item Depending on $\beta U(x)$, $l_0$ and $l_1$ can have either
  positive or negative values. Generally speaking, $l_0$ serves as a
  measure for whether the potential is ``effectively'' attractive
  ($l_0>0$) or repulsive ($l_0<0$).
\item Writing $U(x)=U(x=0)f(x)$, with $f(x=0)=1$ and
  $f(x\rightarrow\infty)\rightarrow 0$, we see that if
  $f(x)$ does not change a sign, then $l_1$ has the same sign as $l_0$.
\item Furthermore, if $|\beta U(x=0)|\ll\ 1$ (weak potential)
  then, from Eqs.~(\ref{eq:l0}) and (\ref{eq:d22}), we readily see
  that $l_0\sim |(\beta U(x=0)|^1$, but $l_1\sim|\beta U(x=0)|^0\gg
  l_0$. Therefore, in this limit,
  \begin{equation}
    d_2\simeq 2l_0l_1, \ \ \ \ {\rm for}\ |\beta U(x)|\ll\ 1.
    \label{eq:d20}
  \end{equation}
\item Finally, we note that $l_0$ and $l_1$ can be associated with the
  {\em quasi-probability distribution}, with statistical weights that
  are given by $w(x)=\exp[-\beta U(x)-1]$ and, thus, may
  also assume negative values. In this statistics,
  $l_0=\int_0^{\infty} w_1(x)dx$ plays the role similar to the
  partition function, while $l_1=\langle x\rangle$ is the average
  displacement.
\end{enumerate}

To check the accuracy of Eq.~(\ref{eq:general}), we consider a
different example of a Gaussian potential
\begin{equation}
  U(x)=Ue^{-(x/x_0)^2}.
  \label{eq:gausspoten}
\end{equation}
We set $x_0=1$, $D=0.01$, and compute $P(x,t)$ from $10^8$ Langevin
dynamics trajectories starting at $x=0$. Fig.~\ref{fig2}(a) shows the
PDF at $t=10^4$ for $U=-0.2$ (black) and $U=+0.2$
(red). Fig.~\ref{fig2}(b) shows the function $Q(x,t)=P(x,t)\exp[+\beta
  U(x)]/G(x,t)-1$ which, supposedly, is well approximated by the
(piecewise) linear form $Q(x,t)\simeq -l_0|x|/2Dt$ since the
quadrupole term is negligibly small. For the examples considered in
the figure, we have $l_0=0.190$ ($U=-0.2$) and $l_0=-0.165$
($U=+0.2$). The dashed red and blue lines in fig.~\ref{fig2}(b) depict
these linear functions and demonstrate that, indeed, they nicely fit
to the function $Q$.

We now switch to the free energy calculation, while keeping only those
contributions that decay either as $\sim 1/t^{1/2}$ or $\sim
1/t$. Taking Eq.~(\ref{eq:general}) and using it in
Eq.~(\ref{eq:energy}) yields the following expression for the
time-dependent energy:
\begin{equation}
  E(t)=2\int_0^{\infty} U(x)Ge^{-\beta
    U(x)}\left[1-\frac{l_0x}{2Dt}+\cdots\right]dx.
  \label{eq:energysm}
\end{equation}
Taking advantage of the fact that contribution to this integral comes
from a finite limited region, we can write that in the limit
$t\rightarrow \infty$, $G\simeq 1/\sqrt{4\pi Dt}=1/Z^*$. This also
allows us to drop the dipole term in the square
parenthesis. Thus~\cite{kb1,kb2},
\begin{eqnarray}
  E(t)&=&\delta E(t)\simeq 2\int_{0}^{\infty} U(x)e^{-\beta
      U(x)}Gd\,x\nonumber\\
  &=& \frac{2\int_{0}^{\infty} U(x)e^{-\beta
      U(x)}dx}{Z^*}+{\cal O}\left(\frac{1}{t^{3/2}}\right),
  \label{eq:energy1}
\end{eqnarray}
which generalizes Eq.~(\ref{eq:free1}) for the energy of a Brownian
particle is a square potential. Eq.~(\ref{eq:energy1}), which can also
be written as
$\lim_{t\rightarrow\infty}Z^*E(t)=\int_{-\infty}^{\infty}
U(x)\exp[-\beta U(x)]dx$, is yet another demonstration that $Z^*$
plays a role similar to a partition function and that the Boltzmann
factor is an infinite invariant density.

For the entropy calculation in the general case, we insert
Eq.~(\ref{eq:general}) into Eq.~(\ref{eq:entropy}), which gives
\begin{widetext}
\begin{eqnarray}
  -TS(t)\simeq 2k_BT\int_0^{\infty}Ge^{-\beta
    U(x)}\left[1-\frac{l_0x}{2Dt}+d_2\frac{x^2-2Dt}{4D^2t^2}\right]
  \left[\ln G-\beta U(x)-\frac{l_0x}{2Dt}-\frac{(l_0x)^2}{8(Dt)^2}
    +d_2\frac{x^2-2Dt}{4D^2t^2}\right]dx,
  \label{eq:entropysm}
\end{eqnarray}
Note that because we are not interested in the contributions to $S$
diminishing faster than $\sim 1/t$, we can (i) truncate the general
solution (\ref{eq:general}) after the quadrupole correction term, (ii)
use the Taylor expansion $\ln[1+\epsilon]\simeq
\epsilon-\epsilon^2/2$, and (iii) omit in the integrand any terms
featuring products of $l_0$ and $d_2$ having dimensionality of $({\rm
  length})^k$ with $k>2$. Rearranging Eq.~(\ref{eq:entropysm}), we
write
\begin{eqnarray}
  -TS(t)&\simeq& 2k_BT\int_0^{\infty}\left\{G\ln G-\beta
  U(x)Ge^{-\beta U(x)}+G\ln G\left[e^{-\beta U(x)}-1\right]-Ge^{-\beta
    U(x)}\frac{xl_0}{2Dt}\left[\ln G -\beta U(x)+1\right]
  \right.\nonumber \\ &+&\left. Ge^{-\beta
    U(x)}\frac{(l_0x)^2}{8(Dt)^2}+d_2\,Ge^{-\beta
    U(x)}\frac{x^2-2Dt}{4D^2t^2}\left[\ln G -\beta
    U(x)+1\right]+\right\}dx.
  \label{eq:entropy2}
\end{eqnarray}
There are six terms in Eq.~(\ref{eq:entropy2}), to be henceforth
denoted by $-TS_i$ ($i=1,\ldots,6$). The first one is simply
\begin{equation}
  -TS_1=-TS_0,
  \label{eq:ts1}
\end{equation}
where $S_0$ is the entropy of the free particle, see
Eq.~(\ref{eq:s0}). The second term is identical to
Eq.~(\ref{eq:energy1}), except for the minus sign; thus,
\begin{equation}
  -TS_2\simeq -E(t)=-2\int_{0}^{\infty} U(x)Ge^{-\beta
    U(x)}dx\simeq-\frac{2\int_{0}^{\infty} U(x)e^{-\beta
    U(x)}dx}{Z^*}.
  \label{eq:ts2}
\end{equation}
In the third term, we explicitly write that
\begin{equation}
  \ln G=-\frac{x^2}{4Dt}-\frac{1}{2}\ln\left(4\pi Dt\right),
  \label{eq:lng}
\end{equation}
which gives
\begin{equation}
  -TS_3=2k_BT\int_0^{\infty}G\left[-\frac{x^2}{4Dt}-\frac{1}{2}\ln\left(4\pi
    Dt\right)\right]\left[e^{-\beta U(x)}-1\right]dx.
  \label{eq:ts3a}
\end{equation}
However, the contribution to this integral is limited to a finite
range, which means that the first term in (\ref{eq:lng}) can be
omitted from (\ref{eq:ts3a}). Further taking the limit
$t\rightarrow\infty$ where $G\rightarrow 1/Z^*$, we arrive at
\begin{equation}
  -TS_3\simeq -\frac{k_BT\ln\left(4\pi
    Dt\right)}{Z^*}\int_0^{\infty}\left[e^{-\beta
        U(x)}-1\right]dx=-\frac{k_BTl_0}{Z^*}\ln\left(4\pi Dt\right).
    \label{eq:ts3}
\end{equation}

For the fourth term in Eq.~(\ref{eq:entropy2}), we substitute
expression (\ref{eq:lng}) for $\ln G$, which gives
\begin{equation}
  -TS_4= -2k_BT\int_0^{\infty}Ge^{-\beta
    U(x)}\frac{xl_0}{2Dt}\left\{-\frac{x^2}{4Dt}+\left[1-\frac{1}{2}\ln
    \left(4\pi
    Dt\right)\right] -\beta U(x)\right\}dx,
  \label{eq:ts4a}
\end{equation}
\end{widetext}
and which we have separated into three terms to be denoted by
$-TS_{4,j}$ ($j=1,2,3$). The third term here 
\begin{equation}
  -TS_{4,3}=2k_BT\int_0^{\infty}Ge^{-\beta U(x)}\frac{xl_0}{2Dt} \beta
  U(x)dx\simeq 0,
  \label{eq:ts43}
\end{equation}
can be neglected because the integral is limited to a finite range. In
the first term
 \begin{equation}
   -TS_{4,1}= 2k_BT\int_0^{\infty}Ge^{-\beta
     U(x)}\frac{xl_0}{2Dt}\frac{x^2}{4Dt}dx,
 \end{equation}
 we notice that most of the contribution to the integral comes from
 the range $x\lesssim \sqrt{Dt}$, which for $t\rightarrow \infty$ is
 much larger than the range of $U(x)$. Therefore, we can set
 $\exp[-\beta U(x)]\simeq 1$ in the integrand, and have
  \begin{equation}
   -TS_{4,1}\simeq
   k_BT\frac{l_0}{4D^2t^2}\int_0^{\infty}x^3Gdx=\frac{2k_BTl_0}{Z^*}.
   \label{eq:ts41}
 \end{equation}
Similarly, the exchange of $\exp[-\beta U(x)]$ with unity in the
second term in Eq.~(\ref{eq:ts4a}) is also allowed, yielding
\begin{widetext}
\begin{eqnarray}
  -TS_{4,2}&=& -2k_BT\int_0^{\infty}Ge^{-\beta
    U(x)}\frac{xl_0}{2Dt}\left[1-\frac{1}{2}\ln\left(4\pi
    Dt\right)\right]dx\simeq
  -k_BT\frac{l_0}{Dt}\left[1-\frac{1}{2}\ln\left(4\pi
    Dt\right)\right]\int_0^{\infty}xGdx\nonumber\\ &=&
  -\frac{2k_BTl_0}{Z^*}\left[1-\frac{1}{2}\ln\left(4\pi
    Dt\right)\right]
  \label{eq:ts42}
\end{eqnarray}
\end{widetext}
Summing Eqs.~(\ref{eq:ts43}), (\ref{eq:ts41}), and (\ref{eq:ts42})
gives
\begin{equation}
  TS_4\simeq \frac{k_BTl_0}{Z^*}\ln\left(4\pi Dt\right)
  \label{eq:ts4}
\end{equation}

For the same reasoning as in the above calculation of fourth entropic
term, it is further permissible to replace $\exp[-\beta U(x)]$ with
unity in the fifth and the sixth terms in
Eq.~(\ref{eq:entropy2}). With this substitution, the fifth term reads
\begin{equation}
  -TS_5\simeq
  2k_BT\int_0^{\infty}G\frac{(l_0x)^2}{8(Dt)^2}dx=k_BT\frac{l_o^2}{4Dt},
\label{eq:ts5}
\end{equation}
and the sixth term is given by
\begin{equation}
  -TS_6\simeq 2k_BT\int_0^{\infty} d_2\,G\frac{x^2-2Dt}{4D^2t^2}
  \left[\ln G -\beta
    U(x)+1\right]dx.
  \label{eq:ts6a}
\end{equation}
In Eq.~(\ref{eq:ts6a}) we identify three terms in the square brackets,
but the contribution of the second one can be neglected because $U(x)$
has a finite range, and the third one vanishes identically. Thus, we
are left with only the first term and, using Eq.~(\ref{eq:lng}) for
$\ln G$, gives
\begin{widetext}
\begin{equation}
  -TS_6\simeq 2k_BT\int_0^{\infty}
  d_2\,G\frac{x^2-2Dt}{4D^2t^2}\left[-\frac{x^2}{4Dt}-\frac{1}{2}\ln\left(4\pi
    Dt\right)\right]dx.
  \label{eq:ts6b}
\end{equation}
The contribution of the second term in square brackets in
Eq.~(\ref{eq:ts6b}) vanishes identically, which leaves us with
\begin{equation}
  -TS_6\simeq -2k_BT\int_0^{\infty}
  d_2\,G\frac{x^2-2Dt}{4D^2t^2}\left(\frac{x^2}{4Dt}\right)=
  -d_2\frac{k_BT}{2Dt}.
  \label{eq:ts6}
\end{equation}

Summing Eqs.~(\ref{eq:ts1}), (\ref{eq:ts2}), (\ref{eq:ts3}),
(\ref{eq:ts4}), (\ref{eq:ts5}), and  (\ref{eq:ts6}) gives
\begin{equation}
     -TS(t)=-TS_0(t)-\frac{2\int_{0}^{\infty} U(x)e^{-\beta
         U(x)}dx}{Z^*}-k_BT\frac{2d_2-l_0^2}{4Dt}.
\label{eq:entropy1}     
\end{equation}
From Eqs.~(\ref{eq:energy1}) and (\ref{eq:entropy1}), together with
Eq.~(\ref{eq:d2}), we finally obtain that the excess free energy
\begin{equation}
    \delta F=E-T(S-S_0)= -k_BT\frac{3l_0^2+4l_0l_1}{4Dt}
    +{\cal O}\left(\frac{1}{t^{3/2}}\right),
  \label{eq:freefinal}
\end{equation}
which generalizes the result of Eq.~(\ref{eq:deltaf1}) suggested in
section~\ref{sec:square} for the square well example. In the limit of
a weak potential, $l_0\ll l_1$ [see Eq.~(\ref{eq:d20})], and
\begin{equation}
  \delta F=E-T(S-S_0)\simeq -k_BT\frac{l_0l_1}{Dt}.
  \label{eq:freefinal0}
\end{equation}
\\
\end{widetext}

\section{Summary and Discussion}
\label{sec:summary}

In this work, we study the problem of a Brownian motion in a
non-confining potential that vanishes at infinity. We start, in
section~\ref{sec:square}, by considering a specific example of
diffusion in a square well potential. In this example, the PDFs, both
inside and outside the well, satisfy the free diffusion equation. We
use the method of images to arrive at
Eqs.~(\ref{eq:p1small})-(\ref{eq:p1large}), where the PDF is expressed
in the form of a multipole expansion with each term decaying
$1/t^{1/2}$ faster than the previous one at asymptotically large
times.  This expansion is generalized in section~\ref{sec:general} to
an arbitrary non-confining (symmetric) potential. The PDF, in the
general case, is given by Eq.~(\ref{eq:general}) with the coefficients
$l_0$ and $d_2$ given by Eqs.~(\ref{eq:l0})-(\ref{eq:d20}).

We use the multipole expansion Eq.~(\ref{eq:general}) to calculate the
Helmholtz free energy of the particle. We arrive at the
Eq.~(\ref{eq:freefinal}) which, to order $\sim 1/t$, is the excess
free energy with respect to that of a free particle. To better
understand this result, it is more instructive to look at the entropy
of the particle, or rather the {\em rate of entropy production}\/,
which can be expressed as a series expansion
\begin{equation}
  \dot{S}=\dot{S}_{\rm leading}+\dot{S}_{\rm 1st}+\dot{S}_{\rm 2nd}+\cdots
  \label{eq:sdot}
\end{equation}
To leading order [see Eq.~(\ref{eq:ts1})], $S_{\rm leading}(t)$ is
equal to the entropy of a freely diffusing particle $S_0(t)$, which is
given by Eq.~(\ref{eq:s0}). The rate of entropy production to this
order is, therefore
\begin{equation}
  \dot{S}_{\rm leading}(t)=\dot{S}_0(t)=\frac{k_B}{t}.
  \label{eq:s0dot}
\end{equation}

The next order term in the asymptotic expression for the entropy is
given by Eq.~(\ref{eq:ts2}), which can be also written as $S_{\rm
  1st}(t)\simeq E(t)/T$. Then, from Eqs.~(\ref{eq:partcoef}) and
(\ref{eq:energy1}), we find that
\begin{equation}
    \dot{S}_{\rm 1st}(t)\simeq-\frac{1}{2t}\frac{E(t)}{T}\sim
    \frac{1}{t^{3/2}}.
  \label{eq:s1dot}
\end{equation}
Notice the correction $\sim 1/t^{3/2}$ to the energy expression
Eq.~(\ref{eq:energy1}). It generates a third order correction to the
$\dot{S}$ that scales as $1/t^{5/2}$ and, therefore, is irrelevant to
the present discussion on the zeroth, first, and second order terms in
the expansion Eq.~(\ref{eq:sdot}). Taking this into account, we note
that due to global energy conservation, the amount of heat which is
transferred to the thermal bath is given by $Q(t)=E(t=0)-E(t)$. The
resulting change in the entropy of the bath is $S_{\rm bath}(t)-S_{\rm
  bath}(t=0)=Q(t)/T=[E(t=0)-E(t)]/T=E(t=0)/T-S_{\rm 1st}(t)$ (plus a
third order correction which is ignored herein). Thus,
\begin{equation}
  \dot{S}_{\rm 1st}(t)+\dot{S}_{\rm bath}(t)=0
  \label{eq:s1dotrev}
\end{equation}
The last result can be interpreted as if the first order correction
describes a reversible process. Of course, the spreading of the
particle is {\em not}\/ a reversible process because $\dot{S}_{\rm
  leading}>0$, i.e., the total entropy in the universe increases, but
the leading correction to this result is simply the negative of the
rate of entropy change in the heat bath. In other words, the 1st
correction term (\ref{eq:s1dotrev}) represents the total change in the
entropy of the particle which is balanced by the change in the entropy
of the bath and, therefore, amounts to no net change in the entropy of
the universe. 

This brings us to the next (2nd) correction to the entropy, which is
given by the sum of the terms in Eqs.~(\ref{eq:ts3}), (\ref{eq:ts4}),
(\ref{eq:ts5}), and (\ref{eq:ts6}). Together, they give
\begin{equation}
  S_{\rm 2nd}=-\frac{\delta F(t)}{T}=k_B\frac{3l_0^2+4l_0l_1}{4Dt}.
  \label{eq:s2nd}
\end{equation}
This is the residual component after the subtraction of the entropy of
a freely spreading particle (zeroth term) and the entropy exchange
with the environment (first term). In contrast to these two terms,
$S_{\rm 2nd}$ depends on the initial distribution of the particle
which, throughout this work, has been assumed to be a delta-function
distribution at the origin. Typically, the lengths $l_0$ and $l_1$
have the same sign [see item \#3 after Eq.~(\ref{eq:d22})], which
means that $S_{\rm 2nd}(t)>0$. Eq.~(\ref{eq:s2nd}) can be interpreted
as if this excess entropy is localized in the region of the potential
and diffuses away with the particle. However, in contrast to $S_{\rm
  1st}$, this component is not transferred to the heat bath. It
diminishes in time at a rate
\begin{equation}
  \dot{S}_{\rm
    2nd}(t)=-\frac{k_B}{t}\left[\frac{3l_0^2+4l_0l_1}{4Dt}\right]
  \sim\frac{1}{t^2},
  \label{eq:s2dot}
\end{equation}
representing a small entropic loss for the universe. This does not
imply a violation of the second law of thermodynamics since we are
only looking at a correction term which is negligible compared to the
entropy gained by the spreading of the particle (\ref{eq:s0dot}).
\\

Acknowledgments: I thank Eli Barkai and Erez Aghion for critical comments on the
manuscript. The support of the Israel Science Foundation (ISF) grant
No.~991/17 is acknowledged.\\


\begin{thebibliography}{99}
\bibitem{comment1} The statement that at asymptotically large times a
  (non-vanishing) function $f(x,t)$ is approximated by another
  function $g(x,t)$ means that for any $x$,
  $[f(x,t)-g(x,t)]\rightarrow 0$, and
  $[f(x,t)-g(x,t)]/f(x,t)\rightarrow 0$ for $t\rightarrow
  \infty$. Formally (mathematically) speaking, this is weak
  convergence.
\bibitem{comment2} A standard derivation of Fick's first law is to
  write the flux $J(x,t)=J^+-J^-$ as the difference between the
  currents $J^+$ and $J^-$ of particles moving to the left and right,
  respectively. Denoting, respectively, by $\tau$ and $d$ the
  characteristic mean free time and mean free path of the particle, we
  can write $J^{\pm}(x,t)=v_0P(x\mp d,t-\tau)$, where the velocity
  $v_0=d/\tau$. The overdamped limit corresponds to $d\rightarrow 0$,
  $\tau\rightarrow 0$, and $dv_0=D/2$, where $D$ is the diffusion
  coefficient. Then expanding $v_0P(x\mp d,t-\tau)=v_0P(x,t)\mp
  D\partial_xP(x,t)/2-d\partial_t P(x,t)+\ldots$, and using this
  expansion to express $J^{\pm}$, Fick law that
  $J=J^+-J^-=-D\partial_x P$ is immediately recovered. In the presence
  of a potential jump, only a fraction $\sigma=\exp[-\beta U]<1$ of
  the current $J^+$ represents particles with enough kinetic energy to
  cross the barrier (based on the overdamped assumption that the
  momentum is distributed according to the equilibrium
  Maxwell-Boltzmann distribution).  Thus, $J^+=\sigma
  v_0P_<(x-d,t-\tau)$, $J^-=v_0P_>(x+d,t-\tau)$, and
  $J=J^+-J^-=v_0(\sigma P_<-P_>)-D\partial_x P$. This means that in
  order to comply with the flux continuity BC (\ref{eq:bcflux}), the
  PDF must also satisfy $\sigma P_<=P_>$ [BC (\ref{eq:bcjump})] at the
  boundary.
\end{thebibliography}
\end{document}